# Contact resistance between two REBCO tapes under load and load-cycles

Jun Lu, Robert Goddard, Ke Han and Seungyong Hahn

National high magnetic field laboratory, Tallahassee, Florida 32310 USA

**Abstract**

No-insulation (NI) REBCO magnets have many advantages. They are self-protecting, therefore do not need quench detection and protection which can be very challenging in a high $T_c$ superconducting magnet. Moreover, by removing insulation and allowing thinner copper stabilizer, NI REBCO magnets have significantly higher engineering current density and higher mechanical strength. On the other hand, NI REBCO magnets have drawbacks of long magnet charging time and high field-ramp-loss. In principle, these drawbacks can be mitigated by managing the turn-to-turn contact resistivity ($R_c$). Evidently the first step toward managing $R_c$ is to establish a reliable method of accurate $R_c$ measurement. In this paper, we present experimental $R_c$ measurements of REBCO tapes as a function of mechanical load up to 144 MPa and load cycles up to 14 times. We found that $R_c$ is in the range of 26-100 μΩ -cm$^2$; it decreases with increasing pressure, and gradually increases with number of load cycles. The results are discussed in the framework of Holm's electric contact theory.

1. Introduction

No insulation (NI) REBCO pancake magnet coils have several advantages over conventional insulated coils [1, 2]. They are self-quench-protecting which eliminates the need for quench detection and protection system that can be very challenging in a high $T_c$ superconducting magnet [3]. NI coil technology also eliminates the volume fraction of insulation and reduces the copper stabilizer volume. This results in magnet coils with very high engineering critical current density and significantly higher engineering mechanical strength that enable a very compact magnet to reach a field well above 30 T.

Some obstacles to overcome in NI magnet technology are long charging delay [4-6] and high ramp losses which is a concern in the operation of cryo-cooled magnets. These issues are directly related to turn-to-turn contact resistivity ($R_c$), and can be mitigated by increasing $R_c$. On the other hand, too high $R_c$ jeopardizes its self-protection ability. Therefore it is highly desirable to have an engineered $R_c$ that minimizes the charging delay and ramp loss while retains reliable self-protection ability. The first step of $R_c$ engineering is to develop a reliable method to accurately measure it.

It is conceivable that $R_c$ varies with variations of conductor surface conditions, such as roughness and native oxide chemistry and thickness. It is also expected to vary with contact pressure and temperature. The turn-to-turn contact pressure in a coil equals the coil radial stress which can be calculated based on coil winding tension, thermal stress and electromagnetic stress for a specific case. A few useful examples are given by [7] where winding stress and thermal stress in a REBCO pancake coil are calculated, and [8] where electromagnetic stress is calculated for a 35 T REBCO magnet. In these cases, the turn-to-turn compressive pressures are below 30 MPa.

$R_c$ of NbTi and $Nb_3Sn$ wires under mechanical load and load cycles have been studied extensively for superconducting cable applications [9-12]. $R_c$ between REBCO tapes has also been studied by a few research groups [13 - 20]. It should be noted that most of REBCO conductors have their REBCO layer grown on a substrate that has an insulating oxide buffer layer. A conductive stabilizer layer is subsequently deposited or laminated surrounding the REBCO/substrate. Therefore the contact resistivity $R_c$ strongly depends on the contact configuration, i.e. whether it is a face-to-face (REBCO-to-REBCO) or face-to-back (REBCO-to-substrate) contact. In mechanical joints for fusion applications [13 - 15] where a minimum $R_c$ is desired, REBCO-to-REBCO configuration is used. In addition, the conductor surfaces are usually ground and cleaned to reduce resistance. In [13 – 15], $R_c$ in order of 0.1 $\mu\Omega$-$cm^2$ were obtained. On the other hand, for turn-to-turn contacts of REBCO pancake coils, conductors are in REBCO-to-substrate configuration. $R_c$ of a SuperPower conductor was measured indirectly from the charging time constant at 77 K [18], and a value of 70 $\mu\Omega$-$cm^2$ was reported. In contrast, a direct measurement of AMSC conductor [20] reported $R_c$ of 0.7-10.6 μΩ-cm$^2$ as a function of pressure. Since the reported $R_c$ values varies considerably, a comprehensive study of $R_c$ in REBCO-to-substrate configuration is still needed. In this paper, we measured $R_c$ in REBCO-to-substrate configuration under load and load cycles at 77 K and below. We discuss our experimental results in the framework of Holm's electrical contact theory.

## 2. Theory of contact resistance

The property of electrical contacts is a special topic of electrical engineering and is generally well understood and termed as Holm's electrical contact theory [21, 22]. Due to surface roughness of a contact, only a very small fraction of surface makes good electrical contacts as shown in figure 1. These contact spots are called asperity spots. Consequently electrical current flow is constricted to these asperity spots, which causes an additional resistance as compared with unrestricted current flow. This is called constriction resistance. In addition, surface contamination such as native oxide film, albeit very thin if in series connection with asperity spots, may make significant contribution to $R_c$ which can be written as [22],

$$R_C = \frac{\rho}{2aN} + \frac{\rho_f d}{\pi a^2 N} \qquad (1)$$

where $\rho$ and $\rho_f$ are resistivity of the contact material (Cu in our case) and the surface contamination film (Cu oxides in our case) respectively, $a$ and $N$ are the average radius and number density (in m$^{-2}$) of asperity spots, and $d$ is the thickness of the oxide film. The first term in equation (1) represents constriction resistance, and the second term is the contribution from the contamination film. This holds true for large numbers of contact situations. Although in case of sub-micron thin films, some deviation from this theory has been observed. [23, 24].

Based on the assumption that the pressure at asperities equals the conductor's plastic flow stress or hardness $H$ [22], radius of asperity spots $a$ can be related to nominal contact pressure $P$ via hardness $H$ of the contact material,

$$H = \frac{P}{\pi a^2 N} \qquad (2)$$

Here $\pi a^2 N$ corresponds to the fraction of the surface that actually made contacts. So equation (1) becomes

$$R_C = \frac{\rho}{2}\sqrt{\frac{\pi H}{PN}} + \frac{\rho_f d H}{P} \qquad (3)$$

In equation (3), the relative contribution of the second term depends on the resistivity and thickness of the film. For example, if the film has been pierced at asperity spots, film thickness is zero, so the second term is zero. If we assume $H$ and $N$ to be independent of load, the $R_c(P)$ curve will be proportional to either $P^{-1/2}$ or $P^{-1}$ depends on which term is dominant.

In practice under increasing mechanical load, one has to consider the change in $H$ due to work hardening especially at cryogenic temperatures, as well as the change in number of asperity spots $N$. Furthermore when oxidized surface surrounds asperity spots, the size of the asperity spots is likely to increase much slower than suggested in equation (3), because asperity spots have to compete with oxides for contact area. When the size of asperity spots can no longer increase with pressure, $R_c(P)$ curve becomes flat.

### 3. Experimental setup

Samples used in this experiment are REBCO conductors made by SuperPower (SCS4050AP). It is 4 mm wide tape with a thickness of about 95 μm including a 20 μm layer of copper stabilizer deposited on each side by electroplating. The nominal critical current in self-field at 77 K is 80 A. The residual resistivity ratio of the copper stabilizer layer is measured on similar SuperPower conductors to be about 50.

For conductor surface morphology and contact cross-section characterization, we use an optical microscope (Olympus BX60M), a scanning electron microscope (Zeiss 1540EsB) and a laser confocal microscope (SLCM, Olympus OLS 3100).

For $R_c$ measurement, a test probe is designed and constructed as shown in figure 2. In this probe, a 25 mm long lap contact in REBCO-to-Substrate configuration is placed on a flat G-10 bottom plate. An alignment tool is used to align two REBCO tapes; the alignment is then checked by a microscope. The centering of the G-10 block on top of the sample is adjusted, so that the thin gaps between the G-10 block and the bottom G-10 plate on either side of the sample is even. Mechanical load is provided by a 6" diameter air cylinder (FABCO-AIR, HP6X4FF) driven by pressurized nitrogen gas whose pressure is measured by a PX209-300G5V pressure transducer (Omega engineering Inc.). The load as a function of gas pressure is calibrated by a 10 kN MTS load cell. The designed maximum load on this probe is 15 kN.

$R_c$ are obtained by linear fits of V-I curves measured between 0 and 15 A. The current source is a PowerTen P63C-51000 (0-5 V, 0 – 1000 A); and voltage meter is a National Instruments SCXI-1125 with a terminal block SCXI-1325 which has a typical noise level of $1 \times 10^{-7}$ V. The voltage tap length is about 75 mm which is about 25 mm outside the contact area on each side (figure 2 (b)).

As-received samples are cut and wiped clean with ethanol and mounted on the sample holder. Then a 24 N load is applied at room temperature to check the centering of the G-10 block under load. This load is maintained during the transportation and cooldown of the probe. For 77 K tests, the sample is immersed in liquid nitrogen. For helium temperature tests, a Cernox temperature sensor is attached to the top of the G-10 bottom plate with GE 7031 varnish. The probe is cooled down to 4.2 K, then let naturally warmup with a typical rate of 10 mK/s. In this case, $R_c$ is measured during the temperature rise automatically by applying +/- 1.00 A current from a Keithley 2400 bipolar DC current source, and measuring voltage with a Keithley 2010 digital multimeter. The error for all $R_c$ measurements is estimated to be less than 0.1 $\mu\Omega$-cm$^2$.

### 4. Results and discussions

a) Morphology of contacting surface

$R_c$ is largely determined by the number and size of contact asperity spots. Therefore it is very important to characterize morphology of contacting surfaces. A low magnification micrograph in figure 3 (a) illustrates the surface morphology of our sample. The longitudinal lines on surface are likely to be scratches made in conductor spooling process. Bright spots seem to be 10-20 μm sized knolls on copper stabilizer layer, as one of them is shown in figure 3(b) in higher magnification. These copper knolls were probably formed during copper electroplating process. The surface roughness $R_a$ as measured by laser confocal microscopy in a 50 x 50 μm² area is 0.196 μm.

The nominal thickness of our samples is 95 μm. But it is expected that the copper stabilizer layer is not very uniform due to the edge effect in copper electroplating process. So the actual thickness profile along the width of the tape is measured by cross-sectional microscopy and shown in figure 4. Thickness obviously varies with three local maximums at the center and both edges of the tape. Consequently when pressed by the flat G-10 block, the center and both edges will take most of the load. This load distribution is confirmed by a pressure-recording film inserted between two contacting surfaces ( 25.4 x 4.0 mm² area) under different nominal pressures at room temperature as shown in figure 5 where darker (red) color indicate higher pressure. Figure 5(a) and 5(b) corresponds to test by two types of films with different sensitivity range (9.7 - 49 MPa and 49 - 128 MPa). It is evident that the pressure is higher at the center and both edges, consistent with the thickness profile shown in figure 4.

In order to directly observe contacts under pressure, a cross-sectional sample is prepared with a 150 MPa nominal pressure which is applied by a pair of G-10 plates compressed by two UNC 8-32 screws tightened with a torque wrench. As shown in figure 6, due to the surface roughness even at such high pressure at room temperature, only a small fraction of area (asperity spots) has intimate contact (labelled in figure 6 as "A").

b) Contact resistance

A typical V-I curve of a REBCO contact measured under a 2.4 MPa pressure at 77 K is shown in figure 7. It is linear except in the region near the critical current of about 80 A (figure 7(a)). When a linear component, obtained by fitting V-I data below 15 A, is subtracted from raw data, the residual as plotted in figure 7(b) reveals non-linearity above 20 A. For this reason, all $R_c$ values presented in this paper are obtained from linear fits of V-I data below 15 A. High linearity below 15 A also indicates that joule heating at the contact is negligibly small.

$R_c$ as a function of pressure is measured up to 144 MPa and shown in figure 8. As expected, $R_c$ decreases with increasing pressure. But it decreases much slower than $P^{-1/2}$ (a simulation with $P^{-1/2}$ is also shown in figure 8) and obviously has a non-zero asymptotic level. In fact, $R_c$ vs. $P$ data can be fitted well with,

$$R_C = \frac{A}{P^{1/2}} + B \qquad (4)$$

Where A = 15.0 and B = 25.4 are fitting parameters. As a comparison, a REBCO-to-substrate soldered lap joint prepared by the method described in [25] has resistivity of 0.95 µΩ-cm² (shown as a dashed line in figure 8), much smaller than the measured $R_c$.

As mentioned above, special care should be taken in analysis of a REBCO-to-substrate contact. Due to the insulating buffer layer, the electrical current must flow from REBCO to the surrounding edges of one conductor before entering the other conductor. The electrical current distribution is very different from the case of REBCO-to-REBCO contact, and can only be simulated numerically [26]. However, since the resistance of a soldered REBCO-to-Substrate joint is only a few percent of $R_c$, it may be concluded that contact resistance as described by Holm's theory is dominant.

$R_c(P)$ curve in figure 8 is significantly flatter than $P^{-1/2}$ or $P^{-1}$ as predicted by equation (3). This discrepancy might be explained by the following. Firstly, the contact copper surface is usually covered with a thin layer of native oxides which are essentially insulating at 77 K [27, 28]. The thickness of surface oxides, or whether it is pierced by asperities, depends on the local pressure. Therefore the proportionality between nominal pressure and the size of asperities as described by equation (2) is no longer strictly valid. Under increasing load, the increase in size of asperity spots is hindered by neighboring oxides covered areas, so $R_c$ decreases much slower than that in equation (3). Secondly, in a REBCO-to-substrate contact, the transport current distribution within the copper stabilizer layer is not uniform and might be pressure dependent. So it is conceivable that $R_c(P)$ curve deviates from equation (3) which is derived assuming uniform current distribution outside the immediate vicinity of the contacting surface. Finally, the copper hardness is expected to increases with load and load cycles due to the work hardening at cryogenic temperatures. This effect will also make $R_c(P)$ curve flatter according to equation (3).

$R_c$ under cyclic load is also measured up to 144 MPa for each cycle and total of 14 cycles. $R_c$ is measured during load cycle number 1, 2, 3, and 14, as shown in figure 9. Small hysteresis in $R_c$ is noticeable for each cycle, where unloading $R_c$ is slightly lower than loading $R_c$ indicative of some degrees of plastic deformation at asperities. It is also noted that after the first load cycle $R_c(P)$ curves can no long be fitted with equation (4). Furthermore under a given load, $R_c$ increases gradually with number of load cycles as shown in figure 10. At 2.4 MPa pressure, $R_c$ seems to level off after only two cycles, while at higher loads $R_c$ is still increasing after 14 cycles. This gradual increase of $R_c$ with number of load cycles have been observed in NbTi CICC [8], but is not very well understood. Based on the results of this experiment, we attribute this effect to cryogenic work hardening of copper under cyclic load.

Similar behavior of $R_c$ versus load and load cycles is observed in three other samples tested. But there is significant variation in $R_c$ from sample to sample. For instance, under 144 MPa pressure $R_c$ spreads from 26 to 60 $\mu\Omega$-cm$^2$ from sample to sample. This appreciable spread is likely due to the variation in sample surface morphology and surface chemistry.

Since most of NI REBCO magnets operate at 4.2 K [1 - 8], measurement of $R_c$ at liquid helium temperatures is also practically important. In addition, $R_c$ temperature dependence may shed light on the nature of the contact, as temperature dependence is material specific. We measured $R_c$ versus $T$ under a nominal contact pressure of 2.4 MPa. As shown in figure 11, $R_c$ is almost constant below 20 K. Above 20 K, $R_c$ increases slowly initially until a rapid upturn with an onset at ~85 K consistent with a sharp increase in longitudinal resistance at $Tc$ of REBCO. The $R_c$ versus $T$ curve below 77 K is generally consistent with the behavior of copper resistivity [29]. This implies that the constriction resistance, which is proportional to copper resistivity, is dominant contribution in the measured $R_c$. However, the ratio of $R_c$ between 77 K and 4.2 K is about 2 as compared to a ratio of 7 measured from Cu stabilizer peeled-off from similar REBCO conductors. This discrepancy may be explained by assuming a $T$ independent contribution of 25 $\mu\Omega$ -cm$^2$. Once this $T$ independent contribution is subtracted, the remaining $R_c$ has a resistance ratio of 7 consistent with copper resistance ratio.

5. **Conclusions**

Turn-to-turn contact resistance in a NI REBCO coil is directly related to the coil charging delay and ramp losses. Contact resistivity $R_c$ of SuperPower REBCO samples are measured at 77 K and 4.2 K under mechanical load between 2.4 and 144 MPa and up to 14 load cycles. We found that $R_c$ is in a range between 26 and 100 $\mu\Omega$ -cm$^2$ and decreases with load, but gradually increases with load cycles. The $R_c$ versus load and load cycle behavior can be understood by Holm's electrical contact theory. In subsequent experiments, we will measure $R_c$ of REBCO conductor coated with different thin films aimed at tailoring $R_c$ for different NI magnet design options.

6. **Acknowledgement**

We thank Dr. Roland Timsit for helpful discussion on interpretation of the results, Brent Jarvis, Vince Toplosky, Jeremy Levitan, Kolby McDaniel and Deven Heard for assistance in experiments. The NHMFL is supported by NSF through NSF-DMR-1157490 and the State of Florida.

Figure captions

Figure 1. A schematic of an electrical contact with electrical current flow from top to bottom. The current flow constriction at the asperity causes additional resistance compared with the bulk conductor. It is called constriction resistance.

Figure 2. (a) a schematic drawing of $R_c$ measurement configuration where contact area is 25 mm long. (b) a picture of experimental probe. The stainless steel push rod with a spherical end provides load. The G-10 block has a concave top surface to match the spherical end of the push rod. The voltage taps are soldered at ~25 mm outside the contact area.

Figure 3. (a) a low magnification picture of a sample surface. Horizontal lines along longitudinal direction are likely to be light scratches made by spooling. Bright spots are small copper knolls 10 – 20 μm in size. (b) a high magnification optical micrograph of the sample surface with a larger copper knoll at the center.

Figure 4. Thickness distribution of a sample along the tape width as measured from cross-sectional micrographs.

Figure 5. Pressure distribution on the sample as measured by pressure-recording films of (a) by film with sensitivity range 9.7-49 MPa, and (b) by film with sensitivity range 49 – 128 MPa. The nominal applied pressure for each measurement is labelled under its picture.

Figure 6. A cross-section of a contact between two REBCO conductors under about 150 MPa pressure. A few asperity spots are evident including one that is labelled 'A'.

Figure 7. A typical V-I curve of a contact. (a) the curve appears to be linear till near sample's self-field 77 K critical current of ~ 80 A. (b) when a linear component is subtracted, V-I curve shows an onset of non-linearity at about 20 A. For this reason $R_c$ values presented in this paper are measured use V-I data below 15 A.

Figure 8. $R_c$ as a function of nominal pressure at 77 K. The solid line is a fit by equation (4) where A = 15.0, B = 25.4. The dashed line represents a simulation of $R_c \propto P^{-1/2}$. As a reference point, the horizontal dashed line is resistivity of a solder joint in REBCO-to-substrate configuration.

Figure 9. $R_c$ measured at $1^{st}$, $2^{nd}$, $3^{rd}$, and $14^{th}$ load cycles up to 144 MPa at 77 K. The lines are guides to the eye. The arrows indicate direction of load change.

Figure 10. Data in figure 8 are replotted as $R_c$ versus number of load cycles at given loads.

Figure 11. Temperature dependence of $R_c$ measured between 4.2 and 100 K and under 2.4 MPa nominal pressure. If a $T$ independent component is subtracted (the dashed line), the $R_c$ ratio between 77 K and 4.2 K is about 7, consistent with copper's $\rho(T)$.

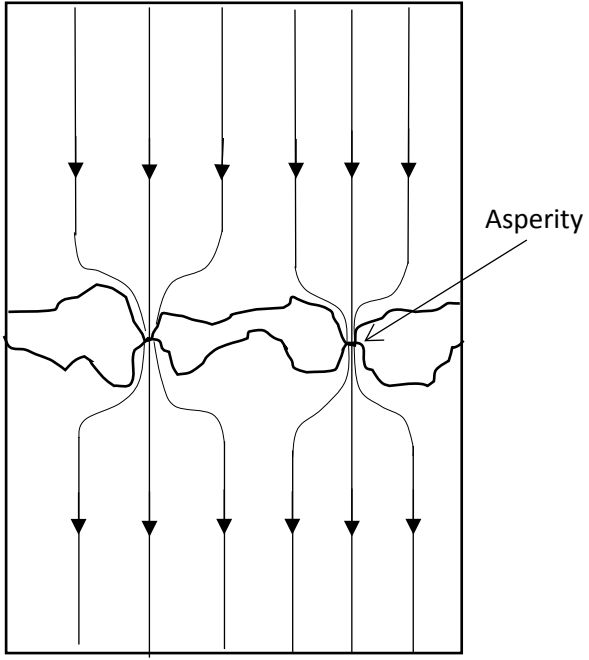

Lu, et al.  Fig. 1

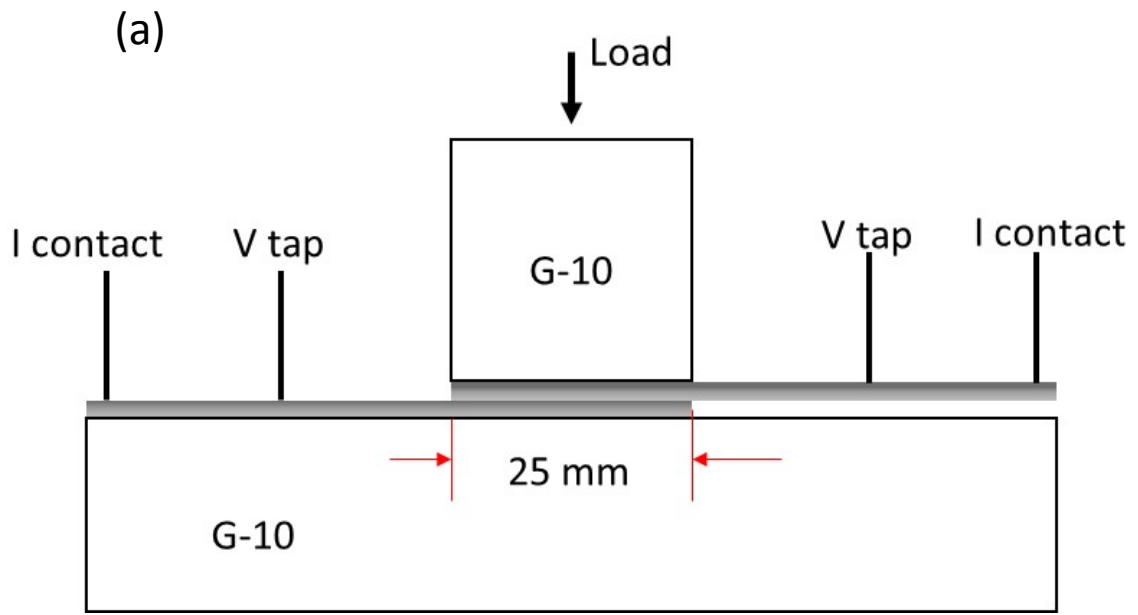

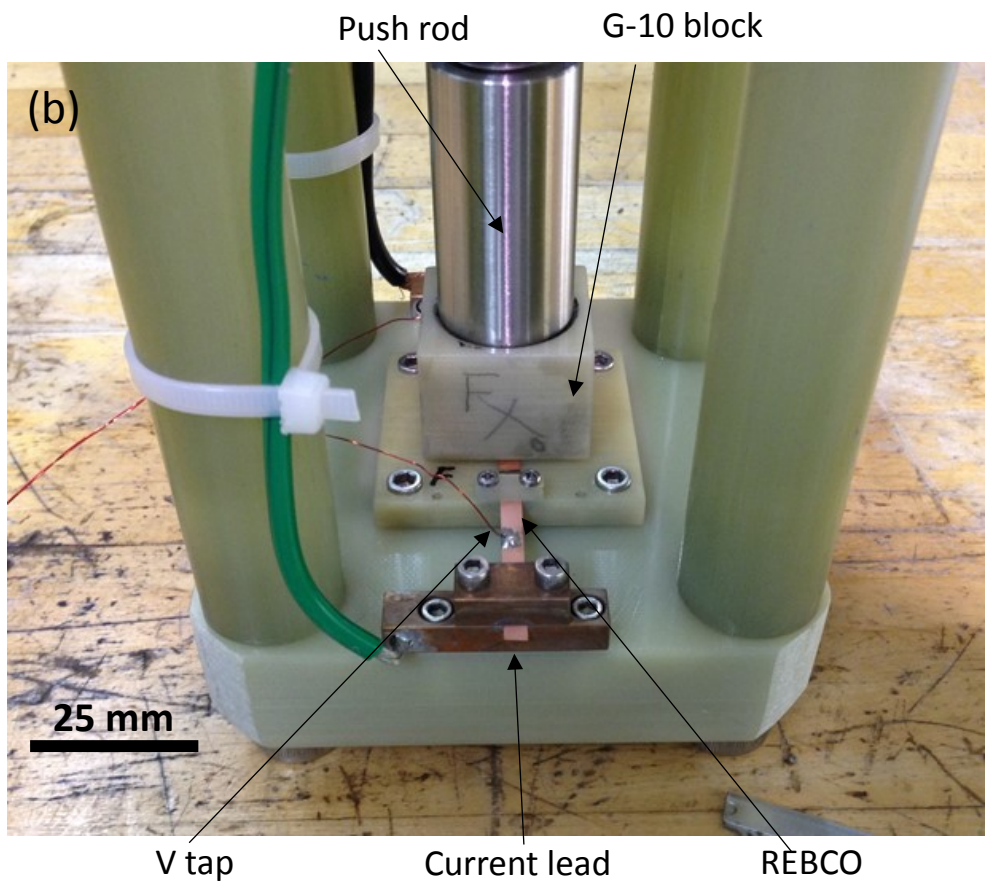

Lu, et al. Fig. 2

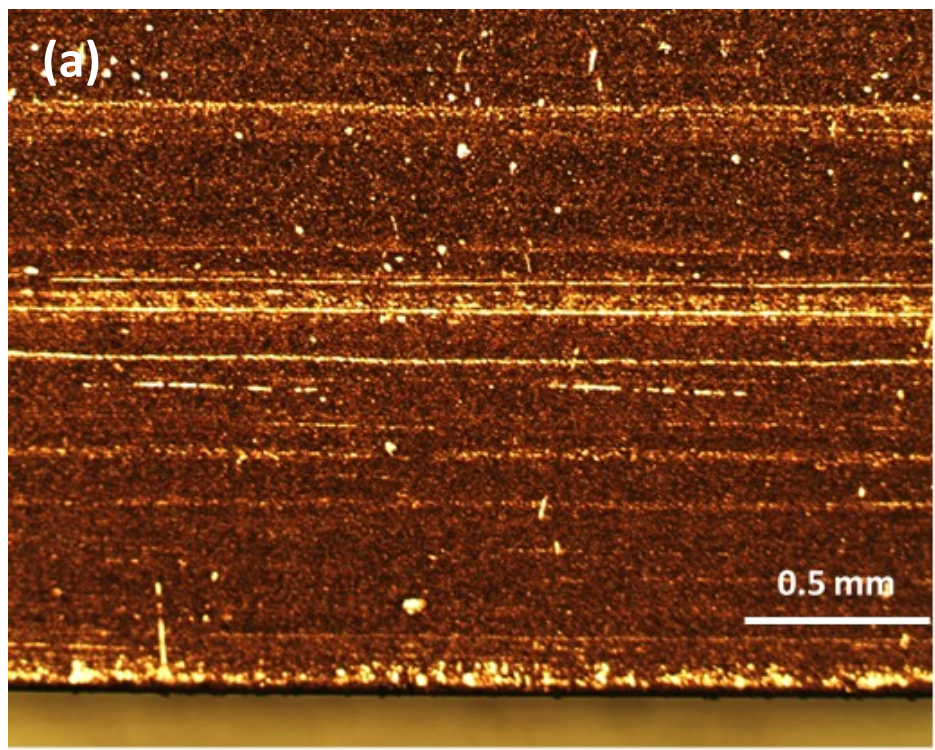

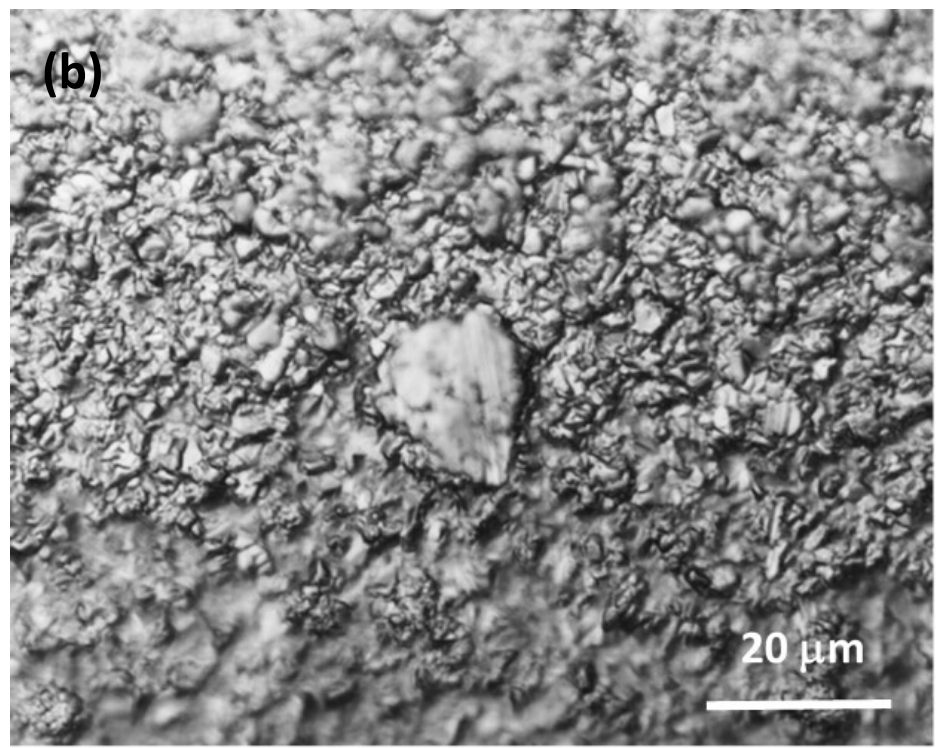

Lu, et al. Fig. 3

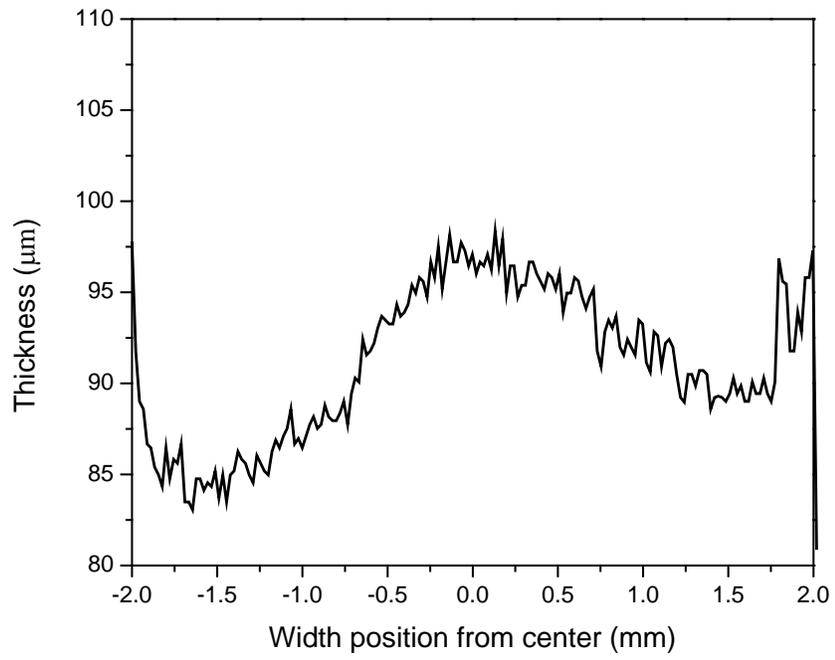

Lu, et al.  Fig. 4

(a)　9.7 – 49 MPa film

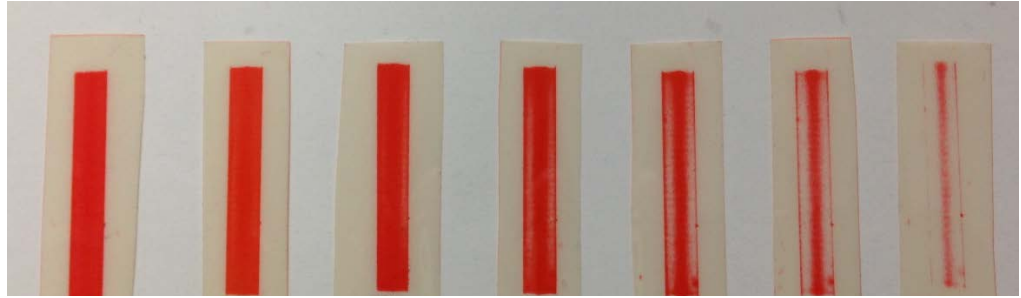

96 MPa　72 MPa　60 MPa　48 MPa　36 MPa　24 MPa　12 MPa

(b)　49 – 128 MPa film

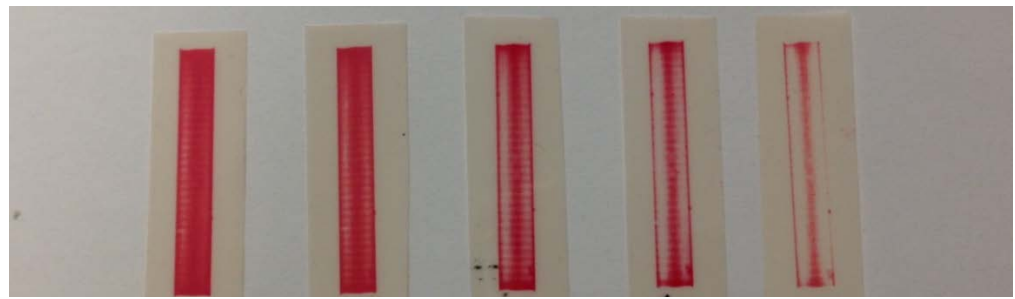

144 MPa　120 MPa　96 MPa　72 MPa　48 MPa

Lu, et al.  Fig. 5

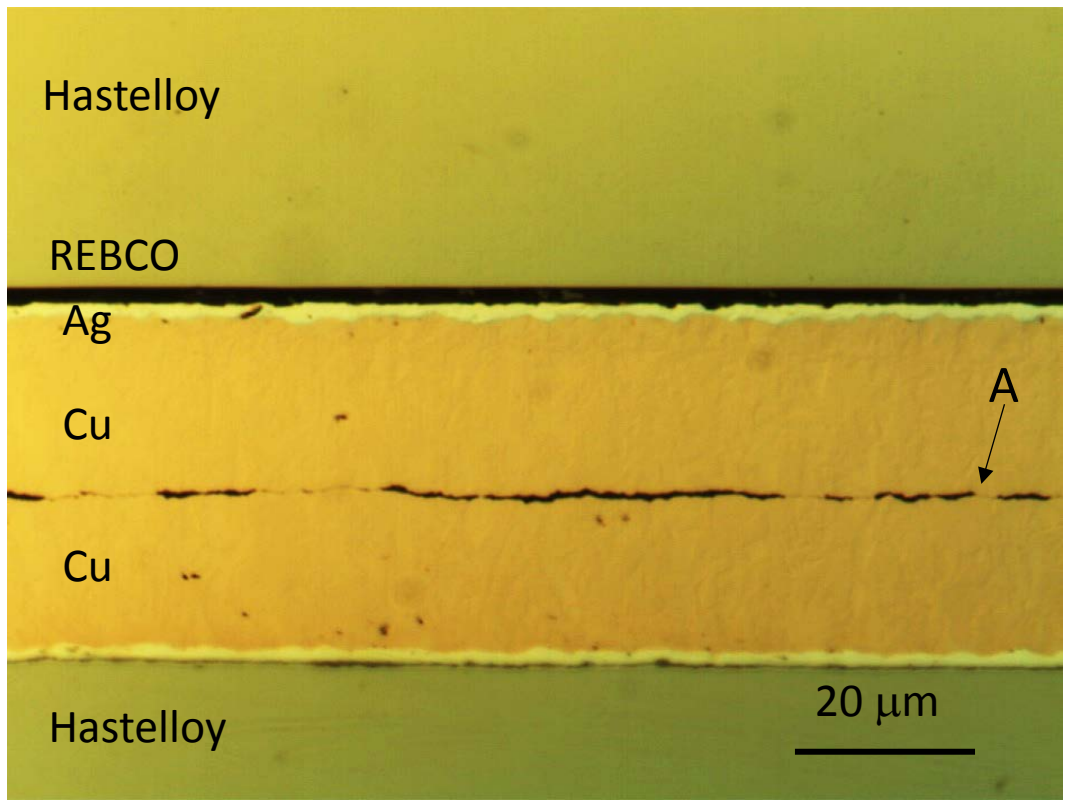

Lu, et al. Fig. 6

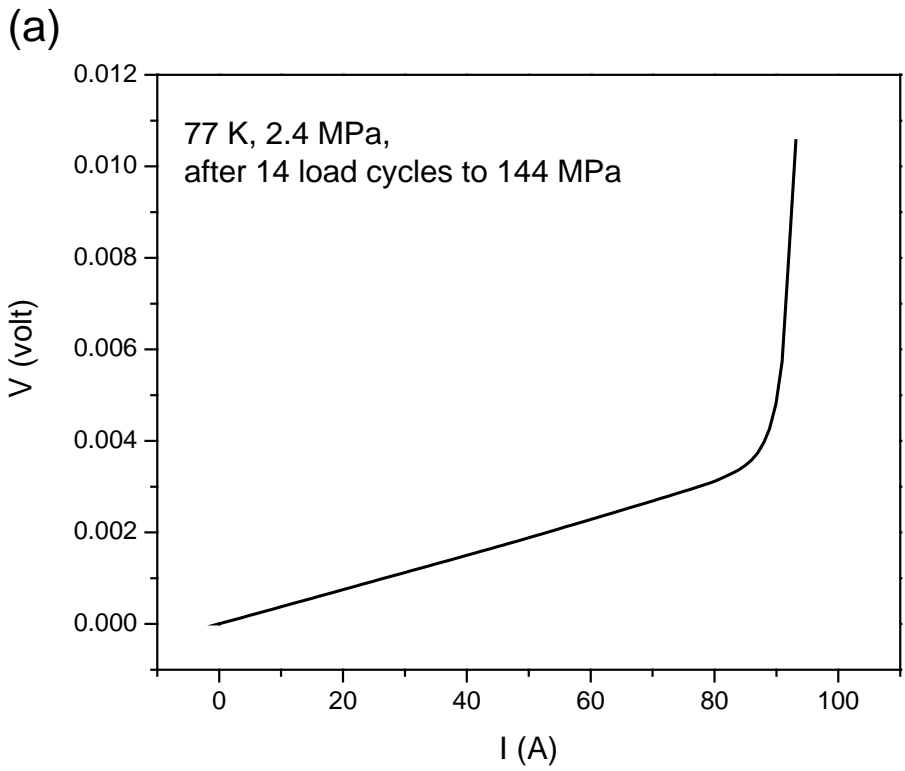

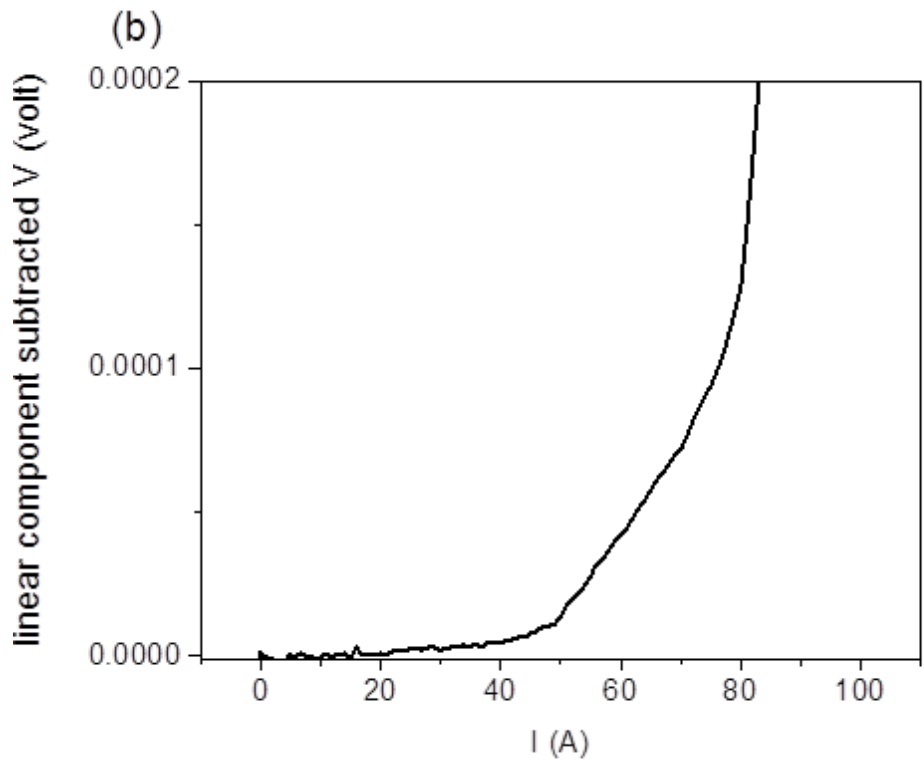

Lu, et al. Fig. 7

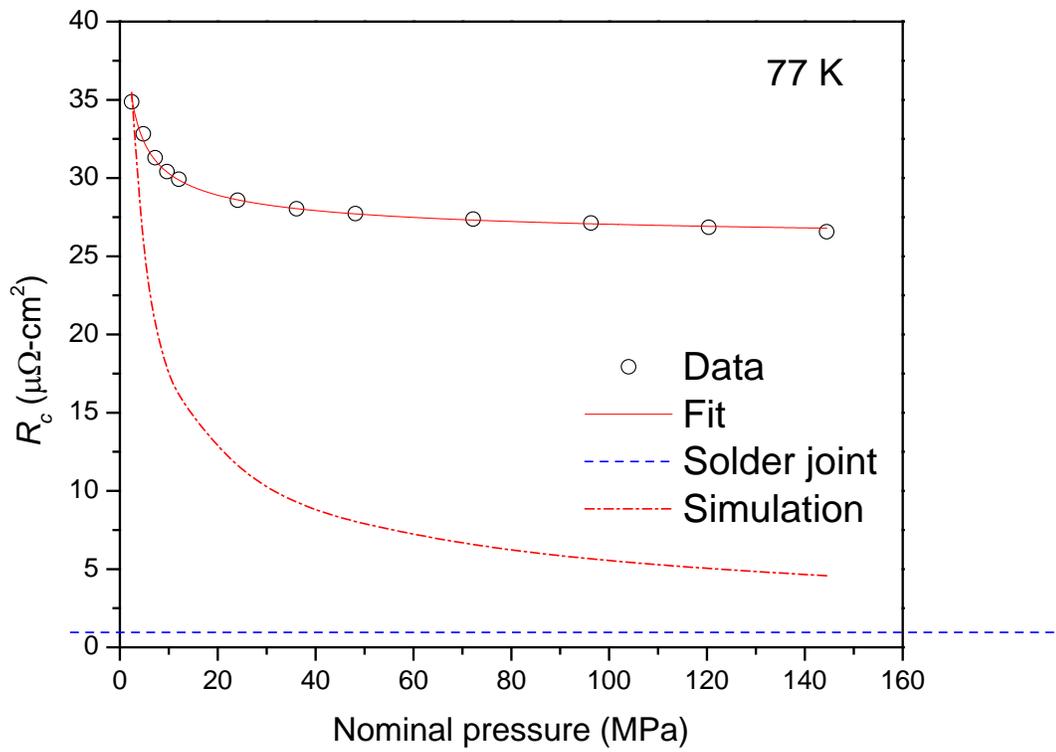

Lu, et al. Fig. 8

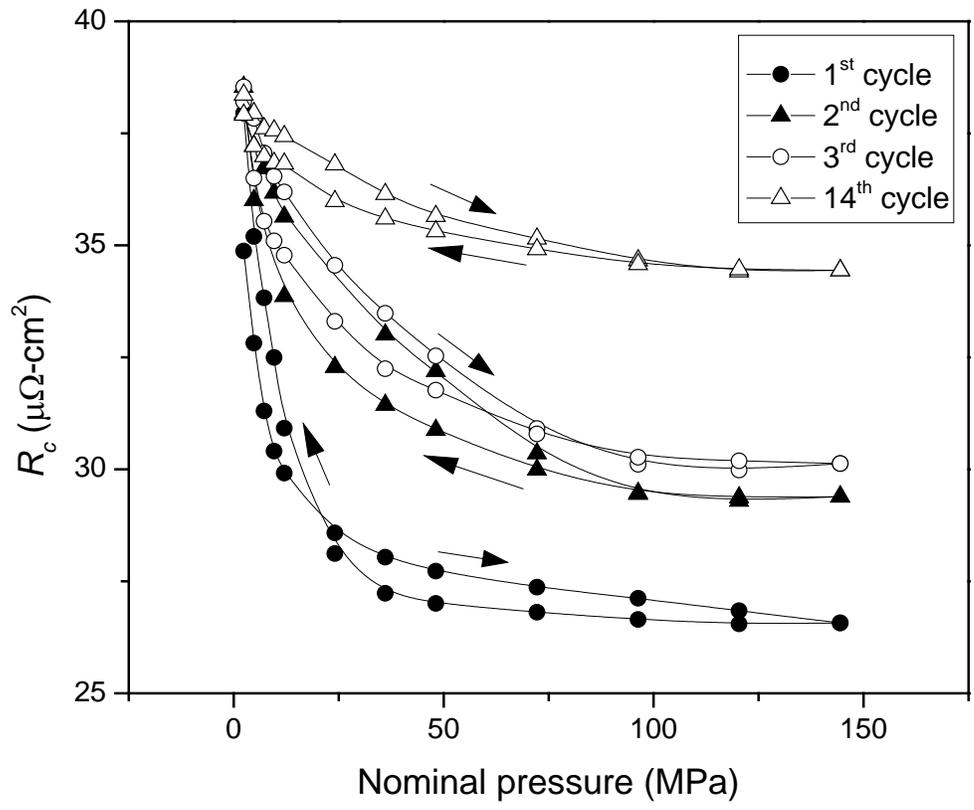

Lu, et al. Fig. 9

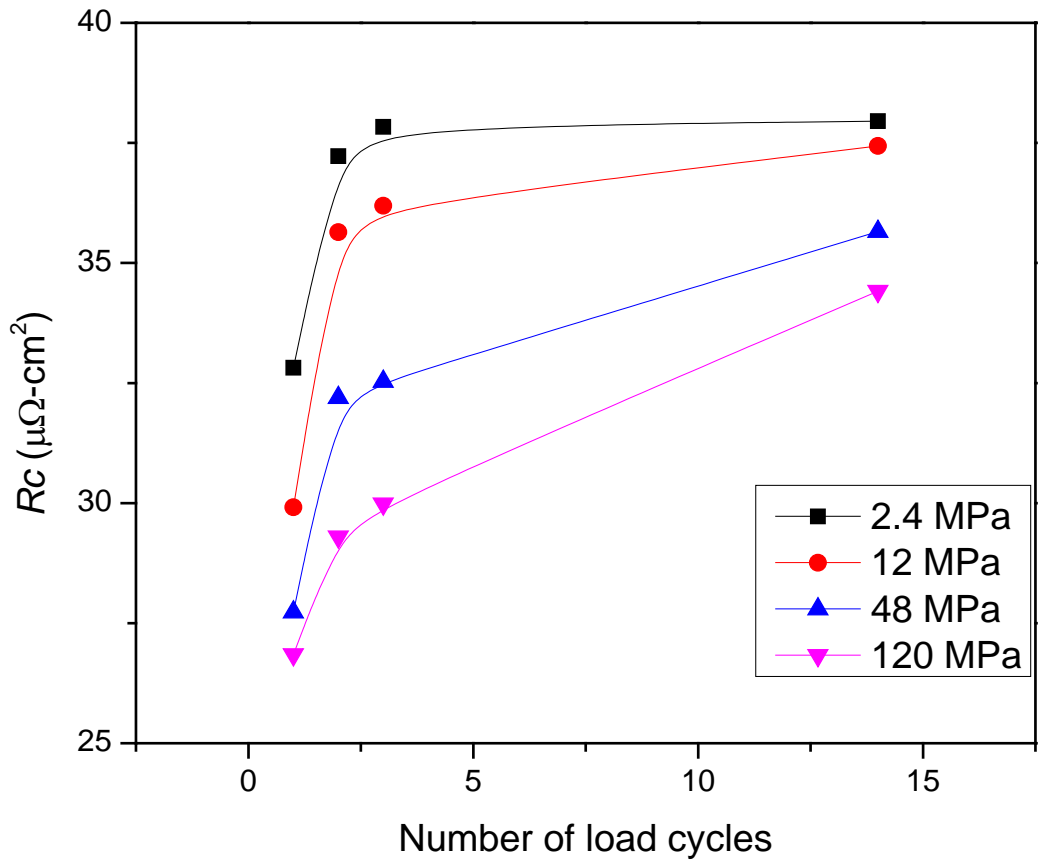

Lu, et al. Fig. 10

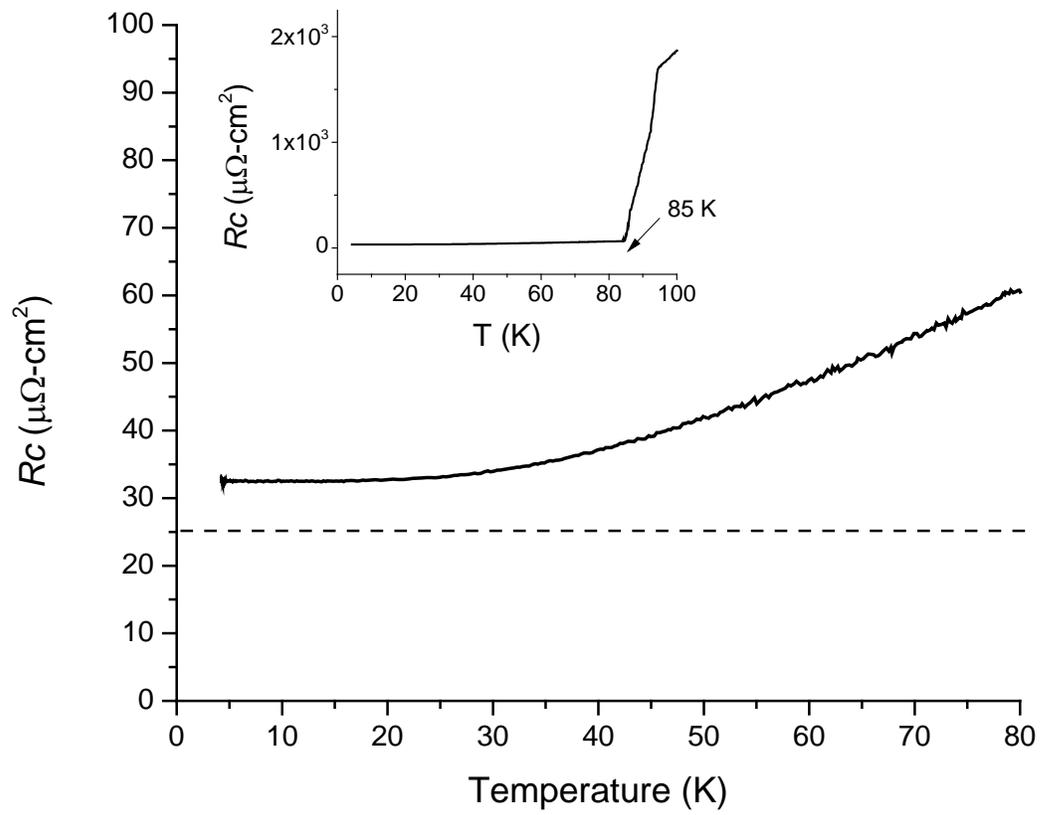

Lu, et al. Fig. 11

24